\documentclass[12pt]{iopart}
\usepackage{graphicx}% Include figure files
\begin{document}

\title{ Angular Momentum of Photon and  Phase Conjugation.}

\author{A.Yu.Okulov}

\address{General Physics Institute \\
of Russian Academy of Sciences \\
Vavilova  str. 38, 119991, Moscow, Russia}

\ead{okulov@kapella.gpi.ru}

\begin{abstract}

Using concept of an $ideal$ phase-conjugating 
mirror we demonstrate that regardless to internal physical 
mechanism the phase-conjugation of singular laser beam 
is accompanied by excitation within mirror of internal waves 
which carry doubled angular momentum in order to match 
angular momentum conservation.
For Brillouin hypersound wavefront-reversal mirror this means that 
each elementary optical 
vortex in a speckle pattern emits acoustical vortex wave with doubled
topological charge.
The exact spatial profiles of light intensity and intensity of hypersound 
in the vicinity of phase singularity are obtained. These $spiral$ profiles 
have a form of a double helix which rotates with the speed of sound. 
The optoacoustic experiment is proposed for visualization of wavefront
reversal of twisted optical beams and tunable twisted sound generation. 

\end{abstract}

\vspace{1cm}

\begin{figure}
\center{\includegraphics[width=0.9\linewidth] {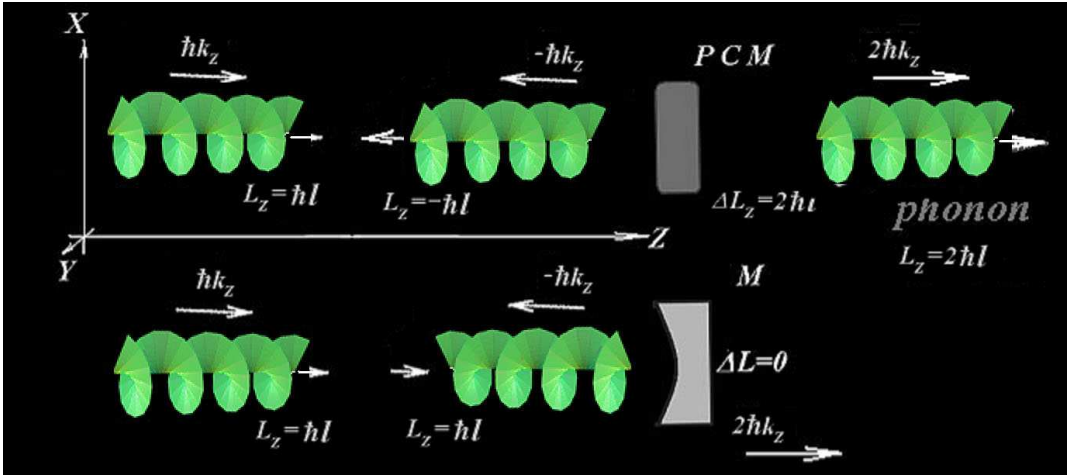}}
\caption{Comparison of conventional mirror ($\bf M$,bottom) and wavefront
 reversal mirror 
 ($\bf PCM$,upper)  from the point of view of 
angular momentum transformation in photon's reflection.
Upper view: The "right" photon with plane polarization and 
helical wavefront strikes $\bf PCM$ and decays to "right" photon, 
moving 
to opposite direction with momentum - $\hbar k_z$. The acoustical 
phonon absorbs 
$translational$ $recoil$ momentum $2 \hbar  k_z$ 
and $rotational$ $recoil$ $\bf OAM $ $2 \hbar $.
Bottom view: The "right" photon with topological 
charge $+1$ strikes the conventional mirror $\bf M$ 
and transforms to "left" photon. In this case the 
 $\bf OAM $ is not changed and $rotational$ $recoil$ is 
absent.}
\label{fig.1}
\end{figure}

The conservation of angular momentum $({\bf AM})$  $  \vec J$ 
stems from  isotropy of space  \cite{Pitaevskii:1982}. 
In contrast to particles with nonzero rest mass $m_o$, the decomposition 
of $ \vec J $  for "spin" 
$  \vec S$  and "orbital" $  \vec L$ parts 
of photon's  ${\bf AM}$ is referred to as 
ambiguous procedure \cite{Pitaevskii:1982,Allen:1992}. 
The spin part $\vec S$ 
is related to polarization, i.e. time-dependent layout of electrical 
$ \vec E$ and magnetic $  \vec B$ fields of the "transverse" light wave. The 
orbital part $({\bf OAM})$ $\vec L$ is associated with helical staircase
wavefront 
\cite{Allen:1992,Berry:1974,Soskin:2001,Barnett:2002}. 
As a matter of fact a purely transverse light waves are abstraction 
because of a 
small but inevitable projections 
of $ \vec E$ and $  \vec B$ on direction of propagation, say $Z$-axis(fig.\ref{fig.1}).
Indeed, the spin-orbital coupling of light occurs \cite{Zeldovich:1992} due to vectorial interplay 
between  longitudinal and transversal components 
of the fields $\vec E $ and $\vec B$. The vectorial solutions 
of Maxwell's equations for propagation of a light spatially localized by 
a waveguide or emitted through finite aperture to free space give a strict  
 relationship between  longitudinal and transversal field 
components  \cite{Jackson:1962,Weinstein:1969}. 
Nevertheless the approximate decomposition in the form 
$  \vec J =  \vec S + \vec L$ proved to be very 
fruitful for small curvatures of light wavefront, i.e. in paraxial wave approximation\cite{Allen:1992}. 

The propagation of light in anisotropic medium changes the polarization and 
historically the spin of photon was observed firstly  in Beth's experiment
 where birefringent $\lambda /2$ plate 
induced the change of photon polarization from circular ($S_z= + \hbar$)
to counter-rotating 
one  ($S_z= - \hbar$) \cite{Beth:1936}. 
The elementary act of photon's spin change accompanied by back action and 
stepwise increase of angular 
momentum of a plate.  The quantum-classical correspondence fulfilled by 
origin of a macroscopically observable classical 
torque $\vec T=\frac {d}{dt} \vec J=[\vec D \times
 \vec E]$  \cite{Allen:1992} , where $|\vec J| \approx { \frac {I} {\omega}}$, 
$\omega$ - is angular frequency and 
$I$ - is intensity of light.

The reflection of circularly polarized photon
from an ideal conventional mirror 
(metal of multilayer dielectric one) 
does not change 
the direction of both spin $\vec S$  and orbital momentum $\vec L$ in 
the laboratory frame and  mechanical torque $\vec T$ on 
such mirror is absent (fig.\ref{fig.1}). This follows both from  
boundary conditions for 
Maxwell equations\cite{Weinstein:1969} and from rotational symmetry of setup
with respect to $Z$-axis\cite{Pitaevskii:1982,Zeldovich:1985}. 

The current communication pays particular attention to conservation 
laws in reflection of photon 
carrying ${\bf OAM}$ $L_z= {\ell}\hbar$ from 
$phase$-$conjugating$ $mirror$ ($\bf PCM$). The discussion 
will be concentrated mainly upon Brillouin wavefront-reversal 
mirror \cite{Zeldovich:1985}.  We will 
demonstrate the hidden anisotropy of SBS-mirror which arise due to excitation 
of internal helical waves, i.e. acoustical vortices, whose existence had been 
proven recently for $Mhz$-range sound \cite{Thomas:2003}. The rotation of 
ultracold cesium atoms \cite{Tabosa:1999} also had been suggested to occur
because of ${\bf OAM}$ transfer due to backward reflection of Laguerre-Gaussian
beam ($\bf LG$) with $0.001$ diffractive efficiency via nondegenerate four-wave 
mixing process. The  
$\bf OAM$ transfer from co-propagating circularly $\bf LG$
to BEC \cite{Wright:1997}had been discussed as well. 

The circularly polarized photon is called "right" when 
projection of the spin $S_z$ upon direction of propagation 
is positive(fig.\ref{fig.1}). 
This happens for example, when photon moves in positive 
direction of $Z-axis$ with 
momentum $\hbar \vec k$ and the fields $ \vec E$ and  $ \vec B$ 
rotate clockwise with
respect to $\vec k$. On the contrary the "left" photon 
have counter-clockwise rotation of polarization and it carries 
spin  $S_z= - \hbar$. 

In reflection from conventional mirror  
the  incident  "right"  photon with  $S_z= + \hbar$ 
moving in positive direction of $Z-axis$ with 
momentum $p_z \approx \hbar |\vec k|$ is transformed in a "left" photon having  
$p_z \approx -\hbar |\vec k|$ and the $same$ spin 
projection $S_z= + \hbar$ . And vice versa, when 
incident "left"(or counter-clockwise) photon strikes mirror, 
the sign of $S_z= - \hbar$ is not changed
in laboratory frame and photon becomes "right". It is not 
surprising because 
setup is isotropic. In this 
situation the only mechanical effect on mirror is light 
pressure \cite{Lebedev:1901,Dunlop:1996}, 
whose major component is normal to mirror surface. 
So the conventional mirror in $normal$ reflection accepts the 
momentum $\Delta p_z \approx 2 \hbar \vec k$ as a single entity  
and does not change both polarization (or spin $\vec S$) and 
orbital momentum $\vec L$.

The situation is changed drastically for $\bf PCM$
\cite{Zeldovich:1985}. The conservation laws determine 
the energy of  
excited acoustical phonon $\hbar \Omega_a$ as a difference of 
energies of incident $\hbar \omega_p$  and reflected 
$\hbar \omega_s$ photons. The 
radiation pressure 
also takes place here because of net momentum transfer 
to sound $p_{phonon}=\Delta p_z \approx 2 \hbar  k_z$ or "recoil". 
The most interesting feature is the conservation of 
$\bf OAM$ $\vec L$. Let us show 
that the wavefront-reversal mirror 
should feel "rotational recoil" 
when vortex beam carrying   ${\bf OAM}$ is reflected with 
ultimate phase conjugation fidelity.
Indeed, because of the 
basic feature of phase-conjugation the retroreflected photon passes 
all states of the incident one in reverse 
sequence \cite{Basov:1980}.This is a so-called $time$ $ reversal$  
property of the   wavefront-reversing mirror.
Consequently the helicoidal phase surfaces of incident photon and 
reflected photon should be perfectly matched (fig.\ref{fig.1}). 
The small mismatch 
of the wavefronts caused by recoil frequency shift and wavenumber 
shift of the order $10^{-5}$ \cite{Basov:1980} 
does not affect the
phase surfaces. 
Because of accurate wavefront's match the ${\bf OAM}$ is 
turned to $180^o$ angle 
and $\bf OAM$ projection 
$L_z= + {\ell} \hbar$ is changed 
to the opposite one $L_z= - {\ell} \hbar$.
As a consequence the  "winding" number 
or topological charge ${\ell}$ does not change the sign 
with respect to propagation direction $\vec k$. 
Thus the conjugated photon with "right" 
$\bf OAM$  remains "right", the photon 
with "left"  $\bf OAM$ remains "left". 

The apparent physical consequence of this fact is the necessity of 
excitation of $internal {\:} wave$ which 
ought to absorb the difference of angular momenta 
$\Delta L_z= + 2 {\ell} \hbar$ before and after 
retroreflection.  Consequently 
this internal wave should have 
singular wavefront unavoidably provided the $\bf PCM$ is ideal.
At this point let us stress again upon remarkable difference 
between $spin$ $\vec S$ and 
$orbital$ $\vec L$ components of angular momentum. The 
electrostrictive nonlinearity 
in isotropic medium is scalar in paraxial approximation 
at least and spin part 
of ${\bf AM}$ is not turned to opposite 
one\cite{Zeldovich:1985}. 
On the contrary, the orbital 
component $\vec L$ or ${\bf OAM}$ does turns, 
because acoustical vortices do exist\cite{Thomas:2003}.
It will be shown below in details by $exact$ $analytical$ $formulas$ 
how acoustical vortices absorb the  ${\bf OAM}$. In fact  
there exist some obstacles in realization of the perfect 
phase-conjugation of elementary optical vortices 
say in the form of $\bf LG$ 
\cite{Starikov:2006}. Nevertheless it will be shown below  
how to overcome this class of difficulties using 
traditional and reliable methods 
of the Brillouin phase-conjugation\cite{Basov:1980}. 

\begin{figure}
\center{\includegraphics[width=0.9\linewidth] {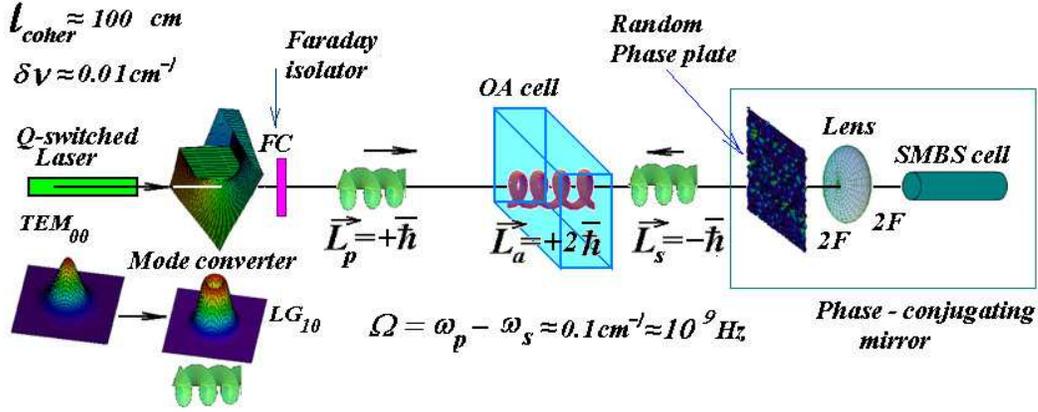}}
\caption{The proposal of experimental setup for twisted hypersound observation. 
The laser field $E_p$ comes through mode converter and  
Faraday cell $\bf FC$ to optoacoustic
 cell $\bf OA$. The transparent random phase plate is 
used for initiation of a high-fidelity  
phase-conjugation in multimode Brillouin-active 
waveguide $\bf SBS$ . $\bf F$-is focal distance 
of a lens.The phase conjugated 
Stockes field $E_s$ interferes with pump field $E_p$ 
inside $\bf OA$ to produce rotating 
spiral interference pattern with transverse size of 
several millimeters. The $twisted$ 
$sound$ $phonons$ excited 
by counter-propagating fields are expected to 
have $\bf OAM $ $L_a = 2 \hbar$ directed to right. 
The angular speed of rotation is equal to acoustical 
frequency $\Omega_a \approx 10^{9} Hz.$ }
\label{fig.2}
\end{figure}

For this purpose let us separate  the region of 
wavefront reversal via traditional phase-conjugating
mechanisms, e.g. Brillouin $\bf PCM$ and optoacoustic
cell  ($\bf OA$) where sound is excited parametrically,
without back action on light(fig.\ref{fig.2}). This geometry
makes possible to visualize inside $\bf OA$ acoustical vortex 
collocated with the optical phase singularity  
in a spatial 
region  having size up to 
several millimeters in transverse dimensions, 
i.e. in $X,Y$-plane\cite{Thomas:2003}.

The acoustical waves inside Brillouin $\bf PCM$ volume 
are highly dissipative, because  
hypersound have a typical relaxation times 
about ${\:}10^{-9} {\:}sec$\cite{Zeldovich:1985}. 
In liquid crystals, which are used for phase-conjugation 
the relaxation times are significantly longer 
- ${\:}10^{-3} {\:}sec$\cite{Zeldovich:1985}. In the 
second  case one might expect to observe the 
macroscopic torque on $ \bf PCM$ for appropriate time scales.
Or else in anisotropic artificial medium alike Veselago's 
lens \cite{Veselago:1968} might appear 
a macroscopic rotational recoil caused by optical torque. 

\begin{figure}
\center{\includegraphics[width=0.7\linewidth] {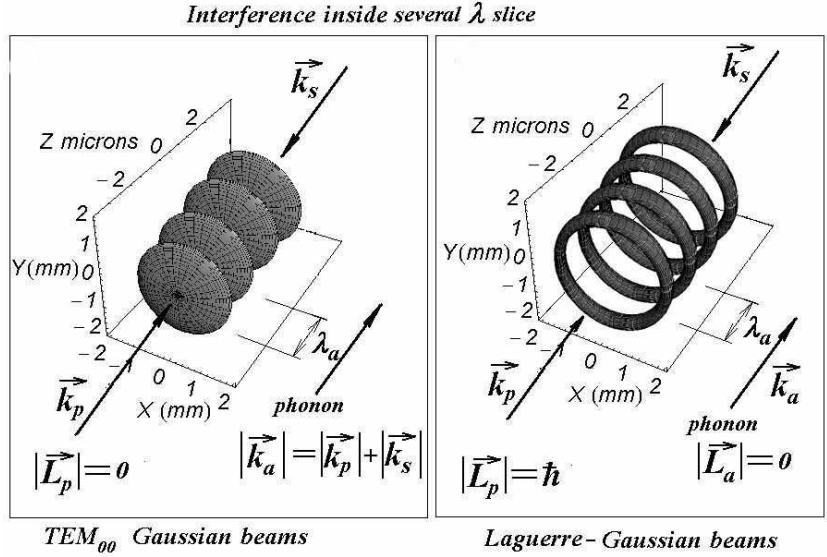}}
\caption{ Interference pattern of incident (pump) Gaussian mode wave $E_p$ with 
 field $E_s$ reflected from conventional mirror. 
Left : incident zeroth-order Gaussian beam forms moving quasi-sinusoidal 
pattern with period $\lambda/2$ due to interference with
reflected beam.Four isosurfaces of interference maxima are shown. 
The direction of pattern motion is changed with the change of the sign of $\Omega$.
The pattern is the same for reflection of $\bf TEM_{oo}$ beam from $\bf M$ and $\bf PCM$.
The period of acoustical grating is equal roughly to a half of the pumping
wavelength in order to satisfy Bragg reflection condition.
Right: $\bf LG$  beam interferes with $\bf LG$ reflected from conventional
mirror.A sequence of toroidal rings occurs. The $\bf OAM $ is not transferred to medium.}
\label{fig.3}
\end{figure}

Consider for definiteness the interaction of the two counter-propagating
 paraxial laser beams inside 
Brillouin active medium. Instead of the quantized field description 
by means of Heisenberg's 
secondary-quantization $\bf \hat \Psi$ - operators 
\cite{Pitaevskii:1982} we choose more 
intuitive approach using  classical counter-propagating optical 
fields $\bf E_p$, $\bf E_s$ and acoustic field $\bf Q$. 
The linearly polarized "pump" field  $\bf E_p$ moves in positive 
direction of $Z-axis$, 
the reflected Stockes  field $\bf E_s$ with the same polarization 
propagates in opposite direction(fig.\ref{fig.3}). 
The acoustic field $\bf Q$ is excited via electrostriction. 
The cylindrical system of  coordinates $(z,r,\phi,t)$ is choosed for
equations below. 
The connection with cartesian coordinates used in 
figures is supposed to be evident. 
The "parabolic" equations of motion are well known\cite{Zeldovich:1985}. 
The "envelope" complex amplitude of 
pump wave $E_p$ moving from left to right (fig.\ref{fig.1}) follows to:

\begin{equation}
\label{pumpwave}
\ {\frac {\partial {{E_p}(z,r,\phi,t )}} {\partial z} }+{\frac {n} {c} }{\frac {\partial {{E_p}}} {\partial t} }+{\frac {i}{2 k_p}} \Delta_{\bot} {E_p} =
{\frac {i \gamma \omega_p} {4 {\rho_0} n c} }Q E_s {\:} ,
\end{equation}

The Stockes  wave  "envelope" $E_s$ moving from right  to  left is controlled by:
\begin{equation}
\label{stockeswave}
\ {\frac {\partial {{E_s}(z,r,\phi,t )}} {\partial z} }-{\frac {n} {c} }{\frac {\partial {{E_s}}} {\partial t} }-{\frac {i}{2 k_s}} \Delta_{\bot} {E_s} = -
{\frac {i \gamma \omega_s} {4 {\rho_0} n c} }E_p {Q }^{\ast},
\end{equation}

The acoustic wave  "envelope" $Q$  moving from left to right obeys to:
\begin{equation}
\label{acouswave}
\ {v_a}{\frac {\partial {{Q}(z,r,\phi,t )}} {\partial z} }+{\frac {\partial {{Q}}} {\partial t} }+{\frac {\Gamma{{Q}}} {2} }=
{\frac {i \gamma {k_a}^2} {16 \pi \Omega_a} }E_p  {E_s }^{\ast},
\end{equation}

where $\gamma=\rho (\partial {\epsilon}/ \partial {\rho})$ - is 
the electrostrictive coupling constant , 
$ {\rho_0} $ - is the density of medium, $n$ - is the index of 
refraction, $c$ - is the speed of light, 
$v_a$ - is a speed of sound\cite{Boyd:1990}. 
The connections between "envelope" complex 
amplitudes $E_p, E_s, Q$ and complete field amplitudes 
$\bf E_p, E_s, Q$ are given by \cite{Zeldovich:1985,Boyd:1990}:

\begin{eqnarray}
\label{pumpwave1}
\ {\bf E_p} = exp{\:} [i( + k_p z-\omega_p t)]{\:} {{E_p}(z,r,\phi,t )}; 
&& \nonumber \\
\ {\bf E_s}=exp{\:} [i(- k_s z-\omega_s t )]{\:} {{E_s}(z,r,\phi,t )}
&& \nonumber \\
\ {\bf Q}=exp{\:} [i( + k_a z-\Omega_a t)]{\:} {{Q}(z,r,\phi,t )}
\end{eqnarray}
The equations  (\ref{pumpwave}-\ref{acouswave} ) are valid within both Brillouin mirror  
and $\bf OA$ (fig.\ref{fig.2}). 
The solution of  (\ref{pumpwave}-\ref{acouswave} )  for $\bf OA$ could be obtained 
under natural physical assumption that amplitudes $\bf E_p$ and $\bf E_s$  are small enough and 
acoustic field $\bf Q$ is excited parametrically, by electrostrictive force in a  right-hand side 
of  (\ref{acouswave}). This assumption enables us to solve  (\ref{pumpwave}-\ref{stockeswave} ) 
 in free-space approximation,  i.e. without right-hand sides. For Cauchi propagation problem with 
first order Laguerre-Gaussian beam as initial condition for  $\bf E_p$ (from left window of $\bf OA$ ) and 
 $\bf E_s$ (from right window of $\bf OA$ ) we have the following $exact$ solutions for pump field $\bf  E_p$ : 

\begin{equation}
\label{pump1}
{{\bf  E_p}(z,r,\phi,t)} \sim {\frac {E^{o}_{p} {\:}  exp {\:} [ {\:}i( + k_p z-\omega_p t)+i{\ell}\phi] {\:}}{ {(1+iz/(k_p  D^2))^2} }} {\:}{r^{\ell}} {\:} 
exp  {\:} [ {\:} - {\frac {r^2}{D^2(1+iz/(k_p  D^2))}} ], 
\end{equation}

and for Stockes field $\bf E_s$ : 
 \begin{equation}
\label{pump2}
{{\bf  E_s}(z,r,\phi,t)} \sim {\frac {E^{o}_{s}  {\:}  exp {\:} [ {\:} i( - k_s z-\omega_s t) +i{\ell}\phi] {\:}}{ {(1+iz/(k_s  D^2))^2} }} {\:}{r^{\ell}} {\:} 
exp  {\:} [ {\:} - {\frac {r^2}{D^2(1+iz/(k_s  D^2))}} ]. 
\end{equation}
The fields $\bf E_p$ and $\bf E_s$ carry angular momentum $\ell \hbar$ per photon\cite{Barnett:2002}, where 
$\ell$ is above mentioned "topological charge" or  "winding number",
$D$ - is diameter of beam "necklace" at full width half-maximum(FWHM), $z$ - is distance passed along $Z-axis$ from beam 
necklace, $r=|\vec r|$ - length of radius vector perpendicular to $Z$, $\phi$-azimuthal angle,
$E^{o}_{s}$ and  $E^{o}_{p}$ - are the maximal electrical field amplitudes in necklace.
The maximally simplified form of the free-space solution choosed
for $\bf LG$ beam \cite{{Okulov:2008}} under assumption that Fresnel 
number for $OA-cell$ is 
large enough $N_f= k_{p,s}D^2/z>>1$ or thickness of $\bf OA$ is 
much shorter than $Rayleigh$ $range$ $k_{p,s}D^2$.

Consider first the interference patterns produced by two counter-propagating fields $\bf E_p$ and $\bf E_s$ 
with equal amplitudes $E^{o}_{p,s}$ and without phase singularities, 
i.e. two zeroth-order Gaussian beams or  $\bf TEM_{oo}$ - beams near 
their overlapping  necklaces:

\begin{equation}
\label{gauss_tem00}
{{\bf E}_{(p,s)}(z,r,\phi,t)} \approx E^{o}_{p,s} {{\:}exp  {\:} [ {\:} -i{\omega_{(p,s)}}t\pm ik_{(p,s)}z-  
\frac {{ r^2}}{D^2(1\pm iz/(k_{(p,s)}  D^2))} ]}  {\:} 
\end{equation}

The isosurfaces of intensity $I_{surface}< 2 |E_0|^2$ as a functions of cylindrical coordinates $(z,r,{\phi},t)$ are the 
solutions of the following $implicit$ equation reminiscent to basic 
course of physical optics: 

\begin{eqnarray}
\label{inter_patt2}
{\bf I}_{isosurface} =|{{\bf E}(z,r,\phi,t)}|^2=|{{\bf E}_p(z,r,\phi,t)}+{{\bf E}_s(z,r,\phi,t)}|^2
&& \nonumber \\
\cong
2 |E^{o}_{p,s}|^2[1+ {\:}cos[  {\:} (\omega_p-\omega_s) t - (k_p+k_s) z]] 
{\:}exp  {\:} [ {\:} - {\frac {2 r^2}{D^2(1+z^2/(k_p^2  D^4))}} ].{\:}{\:}
{\:}{\:}{\:}{\:}
\end{eqnarray}

Apart from familiar interference 
term $cos[ {\:} (\omega_p-\omega_s) t +(k_p+k_s) z]$ which 
describes grating with period $\mathcal {P}= 2 \pi (k_p+k_s)^{-1}$ 
moving along $Z-axis$ with acoustical 
speed $v_a=[(\omega_p-\omega_s)/(k_p+k_s)]$ the Gaussian function 
arosed which modulates the $rolls$ 
of interference pattern in transverse direction. Thus the 
isosurfaces of intensity are pancake-like 
$rotational$ $ellipsoids$ separated by distance $\mathcal {P}$ 
(fig.  \ref{fig.3}).  
Such interference pattern is in a strict agreement with 
the Doppler's mechanism of Brillouin scattering: the pump 
field is being reflected from grating 
which moves with the speed $v_a$. The resulting Doppler 
shift $\omega_p-\omega_s$ is such that optical 
interference pattern perfectly overlaps with spatial profile of acoustic field. 
The left picture of  (fig.  \ref{fig.3}) illustrates this moving
interference pattern for four periods. In conventional picture 
of stimulated Brillouin scattering this 
moving pattern coincides with the moving profile of 
hypersound wave \cite{Zeldovich:1985}. The spatial 
period of this wave $\mathcal {P}=\lambda_a = 2 \pi / (k_{p}+k_{s})$ 
is such that Bragg condition for normal 
reflection from 
moving grating 
(${(\lambda_p)}^{-1}+{(\lambda_s)}^{-1} =  (\lambda_a)^{-1}$) is satisfied. 

Suppose now that first order $\bf LG$ is reflected from conventional,
non phase-conjugating parabolic mirror(fig.  \ref{fig.1}). 
Because orbital angular momentum is not changed in such 
reflection in $laboratory$ frame, the 
helical terms ($\ell \phi$) have identical signs in expressions 
for fields $\bf E_p$ and $\bf E_s$:

\begin{eqnarray}
\label{gauss_lag1}
{{\bf E}_{(p,s)}(z,r,\phi,t)} \approx  E^{o}_{p,s} {{\:} {\:}{r^{\ell}}}
&& \nonumber \\
exp{\:} [ {\:} -i{\omega_{(p,s)}}t \pm ik_{(p,s)}z + i{\ell}\phi] {\:} -
\frac {{ r^2}}{(D^2(1\pm iz/(k_{(p,s)}  D^2)))} ]
\end{eqnarray}

Then 
two counter-propagating  first-order $\bf LG$ produce in $\bf OA$ 
more complicated interference pattern, with a $hole$ on the beam axis. 
As a result instead of sequence of ellipsoids we have 
a sequence of a toroidal rings separated 
by period $\lambda_a=2 \pi (k_p+k_s)^{-1}$ in order to fulfill 
Bragg resonant condition
(fig.  \ref{fig.3}): 

\begin{eqnarray}
\label{inter_patt4}
{\bf I}_{isosurface} =|{{\bf E}(z,r,\phi,t)}|^2=|{{\bf E}_p(z,r,\phi,t)}+{{\bf E}_s(z,r,\phi,t)}|^2 \approx
&& \nonumber \\
{2 |E^{o}_{p,s}|^2[1+ {\:}cos[  {\:} (\omega_p-\omega_s) t - (k_p+k_s) z]] r^{\ell}
{\:}exp  {\:} [ {\:} - {\frac {2 r^2}{D^2(1+z^2/(k_p^2  D^4))}} ]}.{\:}{\:}{\:}
{\:}{\:}{\:}{\:}{\:}{\:}{\:}{\:}{\:}
\end{eqnarray}

Again the interference pattern moves  along $Z-axis$ with 
acoustic speed $v_a$. The direction of 
motion is determined by the sign of 
difference $ (\omega_p-\omega_s)$. For $anti-Stockes$ difference 
of frequencies  the interference pattern moves in 
negative direction of $Z-axis$. 

\begin{figure}
\center{\includegraphics[width=0.7\linewidth] {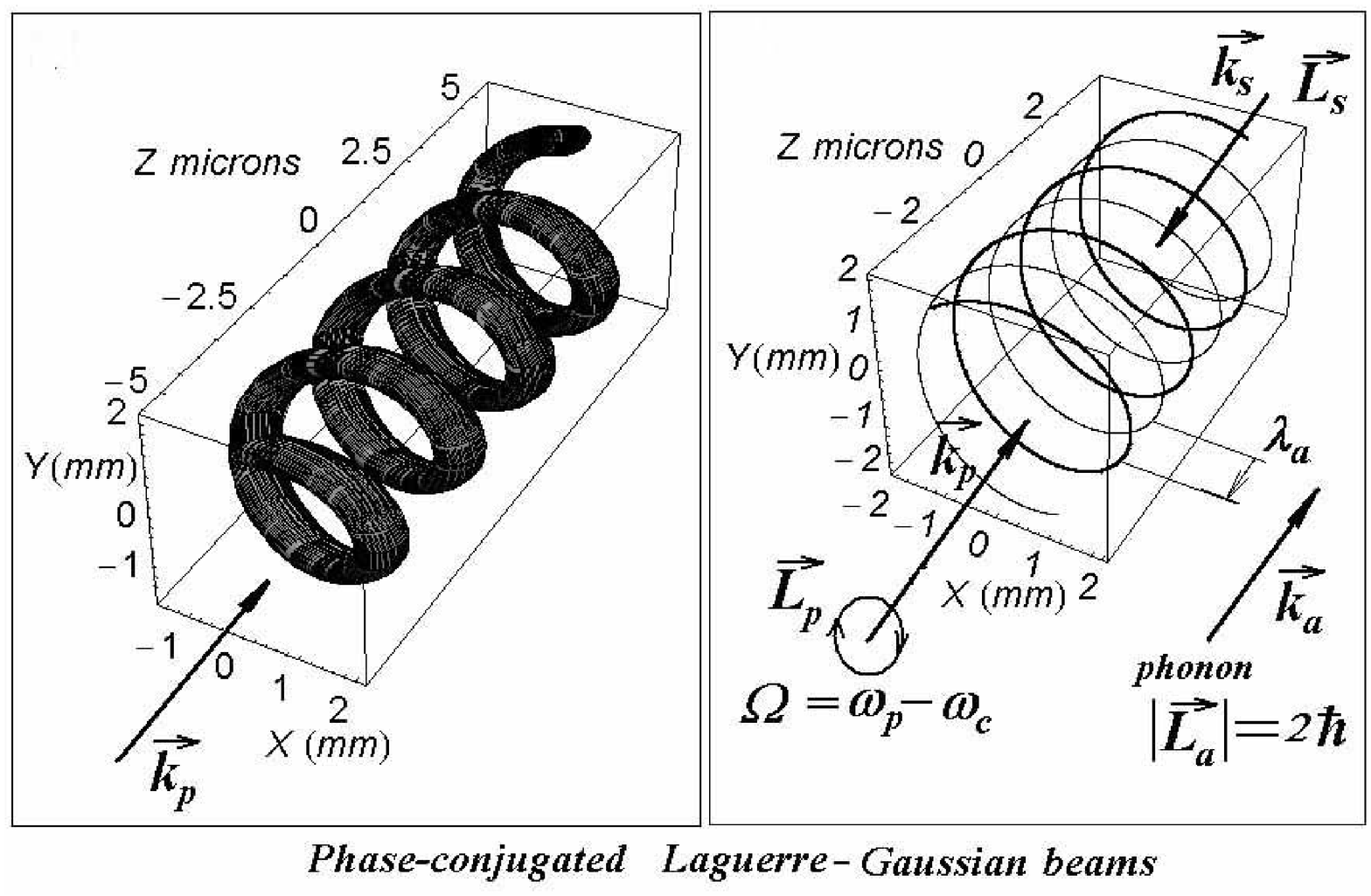}}

\caption{Interference pattern of incident "right" (pump) first order 
$\bf LG$  wave $E_p$ and
phase-conjugated "right" replica  $E_s$. The topological charge is $\ell=+1$. Left: The 
spiral distribution of intensity occurs. The only one string is shown
because of computational power limitations. Right:  Actually the interference pattern have 
the form of a $double$ helix. The right plot depicts the maxima of intensity only. 
The double helix rotates clockwise when the frequency difference is positive(Stockes case).
For anti-Stockes retroreflection the rotation of helix is counter-clockwise. 
The period of acoustical spring is equal roughly to a half of the pumping
wavelength for Brillouin scattering.}
\label{fig.4}
\end{figure}

Phase conjugation of  $\bf LG$ changes the interference pattern drastically. 
Because orbital angular momentum is changed to opposite one in wavefront reversal process, the 
helical terms $\ell \phi$ have $opposite$ $signs$ in expressions for fields $\bf E_p$ and $\bf E_s$ 
and expression for the Stockes field generated by $\bf PC$ - mirror reads: 
\begin{eqnarray}
\label{gauss_lag11}
{{\bf E}_{(p,s)}(z,r,\phi,t)} \approx E^{o}_{p,s} {{\:} {\:}{r^{\ell}}}
&& \nonumber \\
exp{\:} [ {\:} -i{\omega_{(p,s)}}t\pm ik_{(p,s)}z \pm i{\ell}\phi] {\:} -
\frac {{ r^2}}{(D^2(1\pm iz/(k_{(p,s)}  D^2)))} ]
\end{eqnarray}

This mathematical peculiarity, namely  $\pm$ before azimuthal angle $\phi$, 
 follows from the physical fact that after 
retroreflection the ideally phase-conjugated 
wave passes all states of incident wave in reverse 
sequence (\cite{Basov:1980} ). 
As a result the expression for interference pattern is slightly different: 

\begin{eqnarray}
\label{inter_patt1}
{\bf I}_{isosurface} =|{{\bf E}(z,r,\phi,t)}|^2=|{{\bf E}_p(z,r,\phi,t)}+{{\bf E}_s(z,r,\phi,t)}|^2 \approx
&& \nonumber \\
2 |E^{o}_{s,p}|^2[1+ {\:}cos[  {\:} (\omega_p-\omega_s) t - (k_p+k_s) z+2 {l_w}{\:}\phi {\:}]]{\:} r^{2 l_w}
&& \nonumber \\
{\:}exp  {\:} [ {\:} - {\frac {2 r^2}{D^2(1+z^2/(k_p^2  D^4))}} ].
\end{eqnarray}

The self-similar variable ${\:} (\omega_p-\omega_s) t - (k_p+k_s) z$ 
in the argument of $cos$ 
acquires the $doubled$ $azymuthal$ $angle$ $2 {\ell}{\:}\phi {\:}$. 
As a consequence the interference 
pattern changes from sequence of toroidal rings to double 
helix (fig.  \ref{fig.4}). The spiral 
interference pattern has $two$ $maxima$ because azimuthal 
dependence contains $doubled$ 
azimuthal angle $\phi$. 
The spatial 
period of the springs is again automatically adjusted in such a way that one may say 
$generalized$ Bragg condition is satisfied. 
The other drastical feature is that interference pattern $rotates$ with angular velocity equal 
to $acoustical$ $frequency$ $\Omega_a= (\omega_p-\omega_s)$. The linear, translational speed of 
$Z-aligned$ motion 
of a  helix's turns is exactly equal to the sonic speed  $v_a$: 

\begin{equation}
\label{sonic}
\ v_a ={\frac {(\omega_p-\omega_s)}{(k_p+k_s)}}
\end{equation}
 
The similar double-helix structures
were reported recently for a Wigner crystals in a dusty plasma \cite{Tsytovich:2007}. 

The singular acoustical fields in $Mhz$-range were 
generated and measured \cite{Thomas:2003}. The helical acoustical wavefronts 
 were recorded. 
Let us to obtain the exact expression for 
spatial distribution of the sound intensity near optical phase singularity in 
Brillouin medium using equation for acoustical field ${\bf Q}_{twisted}$.
The expression for intensity of hypersound field $I_{sound}$ reads 
immediately from 
the general expressions for 
electrostrictive nonlinearity \cite{Zeldovich:1985,Boyd:1990}:  
\begin{eqnarray}
\label{twistedacous}
\ {\bf I}_{sound} \approx |{{\bf E}(z,r,\phi,t)}|^2=  |E^{o}_{p,s}|^2 
[1+ {\:}cos[  {\:} (\omega_p-\omega_s) t - (k_p+k_s) z+2 {\ell}{\:}\phi {\:}]] 
&& \nonumber \\
{\:}r^{2 \ell} {\:} exp  {\:} [ {\:} - {\frac {2 r^2}{D^2(1+z^2/(k_p^2  D^4))}} ].
\end{eqnarray}

We designated this acoustic vortex field as ${\bf Q}_{twisted}$ having in mind 
helical distribution of intensity (fig.  \ref{fig.4}). It is clear that 
such vortex possesses $doubled$ angular momentum, as it seen from doubled azimuthal angle $2 {\ell}\phi {\:}$ 
in self-similar argument of $cos$ in (\ref{twistedacous}) . 
In addition the complex acoustic envelope $Q_{twisted}$ could be derived in steady-state regime 
due to the fact of the strong dumping of acoustical field. The lifetime for hypersound wave $\Gamma^{-1}$ 
is of the order of the several nanoseconds for typical liquid and gaseous media\cite{{Zeldovich:1985}}.
Thus expression for acoustical envelope ${ Q}_{twisted}$ is obtained in steady-state from (\ref{acouswave}) 
using (\ref{gauss_lag1}) :
\begin{equation}
\label{twisted_ac_env}
\ Q_{twisted} \approx E_p  {E_s }^{\ast} \approx{\:} {exp{\:}[  {\:}+i2 {\ell}{\:}\phi {\:}]}
{\:}r^{2 \ell} {\:} exp  {\:} [ {\:} - {\frac {2 r^2}{D^2(1+z^2/(k_p^2  D^4))}} ].
\end{equation}

Evidently the envelope of acoustic wave ${ Q}_{twisted}$ has helical wavefront with doubled
topological charge $2 \ell$. The twisted 
spatiotemporal acoustic field ${  {\bf Q}}_{twisted}$ has the following form: 

\begin{eqnarray}
\label{twisted_ac_field}
\ {\bf Q}_{twisted} \approx  {\bf E}_p  { {\bf E}_s }^{\ast} \approx{\:} {exp{\:}[  {\:}i (\omega_p-\omega_s) t - i(k_p+k_s) z+i2 {\ell}{\:}\phi {\:}]}
&& \nonumber \\
{\:}r^{2 \ell} {\:} exp  {\:} [ {\:} - {\frac {2 r^2}{D^2(1+z^2/(k_p^2  D^4))}} ].
\end{eqnarray}

The phase dislocation of acoustic wave rotates with acoustical frequency $\Omega_a=\omega_p-\omega_s$.  The speed of translational 
motion in $Z-direction$ of rotating turns of an acoustical spring is exactly equal to the speed of sound  $v_a$. The rotating 
spring could be visualized by currently available experimental 
tools \cite{Thomas:2003}, because the transverse size of beam necklace 
could be easily changed e.g. by additional lenses.

The previous exact expressions  (\ref{inter_patt1}, \ref{twistedacous}, \ref{twisted_ac_field}) are based upon 
elementary exact solutions of the $parabolic$ 
wave equations in the form of the first  order ${\bf LG}$. This simplification became possible due to 
geometrical separation of the $\bf OA$ with weak interaction of counter-propagating beams from 
${\bf PCM}$ where strong interaction of  the optical fields $\bf E_p$ and $\bf E_s$ takes place. 
The straightforward generalization is to be made 
taking into account the $elongated$ $geometry$ of phase singularities inside 
speckle pattern within SBS-mirror volume (fig.  \ref{fig.2}) \cite{{Zeldovich:1985}}. The expression for the optical fields near the 
phase singularity  with topological charge ${\ell}$ could be 
generalized in the following form:

\begin{equation}
\label{gauss_lag5}
\ {{\bf E}_{(p,s)}(z,r,\phi,t)}= E^{o}_{p,s} {{\:} {\:}{r^{\ell}} {\:} exp  {\:} [ {\:} -i{\omega_{(p,s)}}t\pm ik_{(p,s)}z \pm i{\ell}\phi] {\:} ] f(r,z)},
\end{equation}
where $f(r,z)$ - is a smooth  function elongated in $Z$-direction. 
The inequality 
of the forward  ${\bf E}_{p}$ and  backward ${\bf E}_{s}$ fields 
amplitudes does not affect qualitatively 
the helical interference patterns (fig.  \ref{fig.4}) in the regime of weak 
saturation, because the Brillouin wavefront reversal mirrors 
with random phase plate 
have sufficiently
high (approximately 0.9 ) phase-conjugating fidelity\cite{Basov:1980}:

\begin{equation}
\label{correl}
\ K= {\frac  {{ |\int {{\bf E}_{p} {\bf E^{\ast}}_{s}} d \vec r |^2}} {{(\int |{\bf E}_{p}|^2 d \vec r )}{(\int |{\bf E}_{s}|^2 d \vec r )}}},
\end{equation}

\begin{figure}
\center{\includegraphics[width=0.9\linewidth] {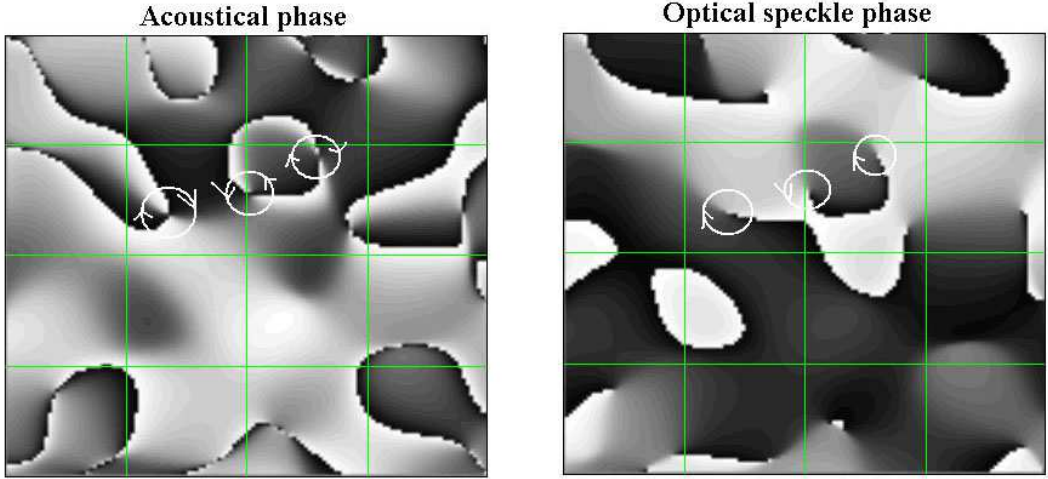}}
\caption{Comparison of acoustical (left) and optical $speckle$ (right) fields 
inside Brillouin mirror pumped by multimode random phase field. 
The optical vortices on right panel are 
designated by white circles with one arrow. The phase is changed 
from $0$ to $\pm 2 \pi$ in  motion around $optical$ phase singularity. 
The collocated acoustical vortices 
on the left panel are shown by white circles with two arrows. The phase is changed 
from $0$ to $\pm 4 \pi$ in  motion around $acoustical$ phase singularity. The transverse 
$X$ , $Y$ scale is $100 \times 100$ $\mu$m. }
\label{fig.5}
\end{figure}

The experimentally verified procedure utilizes the thin transparent glass plates with 
chaotic phase modulation (fig. \ref{fig.2}) which have transverse correlation length of 
about tens of microns. Such geometry looks promising for 
realization of the phase-conjugation of  ${\bf LG}$ and to overcoming a difficulties 
reported earlier \cite{Starikov:2006}. Because of high degree of 
phase-conjugation fidelity of the  forward  ${\bf E}_{p}$ and  backward  ${\bf E}_{s}$ fields  
one may  consider the transverse distributions of pump and Stockes fields near entrance window 
of $\bf PCM$ as almost identical. Thus the above 
expressions (\ref{inter_patt1}, \ref{twistedacous}, \ref{twisted_ac_field})  for spring interference patterns 
might be valid for speckle fields as well. 
This enables us to plot transverse distributions of the  optical and acoustical phases in output planes 
of ${\bf PC}$-mirror and ${\bf OA}$ in a similar way, where phase-conjugated Stockes field $\bf E_s$  is in a good correlation 
with the pump field $\bf E_p$. Choosing the $\bf 8 \times  8$ random plane waves superposition 
for  $\bf E_p$ and $\bf E_s \approx \bf {E^{\ast}}_p$  we obtained using (\ref{twisted_ac_env}) the 
transverse distributions of the phase of the envelope of 
optical speckle field $arg ({ E_p}(\vec r_{\bot})) $ and for envelope 
of a field of collocated acoustical vortices 
$arg ({ Q_{twisted}}(\vec r_{\bot})) $. The acoustical vortices have 
doubled  topological charges (fig.  \ref{fig.5}). 

In summary we demonstrated that the wavefront - reversal mirrors are chiral optical antennas.
The elementary consideration of conservation 
laws including conservation of $\bf AM$ shows the existence of twisted rotating structures inside 
 an $ideal$  phase-conjugator.For stimulated Brillouin 
scattering mirror the rotating 
spiral interference pattern modulates the dielectric 
permittivity $\bf {\epsilon}$ via electrostriction.
The dynamical equations give the exact expressions for 
sound intensity inside acoustical phase singularity 
which is $collocated$ with optical phase singularity. 
It is worth to mention the model presented above does not take into 
account the several essential features of  Brillouin $\bf PCM$,
e.g. longitudinal dependence of optical fields and corresponding spatial 
mismatch of their amplitudes.
The reflection of the both elementary optical vortices like $\bf LG$ and 
speckle  fields as well is accompanied by excitation inside SBS 
wavefront-reversal mirror of $sonic$  vortices with doubled topological charge. 
Internal helical waves and spiral modulation of 
dielectric permittivity induce local anysotropy 
inside the phase-cojugating mirror and 
forces the exchange of angular momentum. 
The new experimental geometry is proposed (fig.  \ref{fig.2}) 
in order to observe 
parametric excitation of an isolated acoustical vortices 
inside $\bf {OA}$. 
Interference pattern near each phase singularity in a speckle pattern 
rotates clockwise with angular speed 
$\Omega=\omega_p-\omega_s$ regardless to the sign of
spirality of the interference spring. The rotation changes direction 
to counter-clockwise for all singularities in a speckle for anti-Stockes 
frequency of retroreflected wave. The angular speed $\Omega$ depends also 
on the physical mechanism of wavefront reversal. It 
could span from units of $Hz$ for 
the photorefractive crystals to $terahertz$  range 
for Raman phase-conjugators\cite{Zeldovich:1985}.

The connection 
of the angular momenta of the incident and reflected 
photons seems to be valid for any other 
nonlinear ${\bf PCM}$ 
including photorefractive one. In order to transfer $\bf OAM$ the 
rotating intensity 
springs should excite the helical waves intrinsic to 
given type of phase conjugating mirror. The peculiarities of photorefractive 
media, e.g. nonstationary vortex reflection, screening,vortex splitting
and "disappearance of nonlinearity" \cite{Mamaev:1996} also 
deserves special attention. The successful wavefront reversal of complex images 
obtained else the first experiments with photorefractive 
phase conjugators \cite{Feinberg:1981} is an indirect evidence 
for existence of internal helical waves in the volume of photorefractive 
 $\bf PCM$  and other phase-conjugators alike 
liquid-crystal light valves.  

The correspondence between formulas for classical 
fields $\bf E_p$, $\bf E_s$ and $\bf Q$ and quantum field description 
by means of Heisenberg's secondary-quantization $\bf \hat \Psi$ - operators 
 \cite{Pitaevskii:1982} will be given elsewhere. Briefly in quantum picture of 
ideal phase conjugator (fig.\ref{fig.1}) each incident 
photon with orbital angular momentum $L_z=\ell \hbar$ decays to 
reflected photon with opposite  $L_z= -\ell \hbar$ projection on propagation 
axis  and a quantum of internal wave with 
doubled $\bf OAM$  ${L_z}^a=2 \ell \hbar$.

\end{document}